\documentclass[letter]{aa}  
\usepackage{graphicx}
\usepackage{txfonts}
 \usepackage{graphicx}
 \usepackage[normalem]{ulem}
 \useunder{\uline}{\ul}{}
 \usepackage{lscape}
 \usepackage{xcolor}
 \usepackage{hyperref}

\begin{document} 
\title{ The First Ground Level Enhancement of Solar Cycle 25 on 28 October 2021}

  \author{
          A. Papaioannou \inst{1}
  \and
          A. Kouloumvakos \inst{2}
  \and
          A. Mishev\inst{3}
  \and
          R. Vainio \inst{4}
  \and
         I. Usoskin\inst{3}
  \and
          K. Herbst \inst{5}
  \and
          A. P. Rouillard \inst{2}
  \and
          A. Anastasiadis \inst{1}
           \and
          J. Gieseler \inst{4}
           \and
          R. Wimmer-Schweingruber \inst{5}
           \and
          P. K\"{u}hl \inst{5}
   }    

  \institute{
              Institute for Astronomy, Astrophysics, Space Applications and Remote Sensing (IAASARS), National Observatory of Athens, I. Metaxa \& Vas. Pavlou St., 15236 Penteli, Greece\\ \email{atpapaio@astro.noa.gr}
  \and
              IRAP, Université Toulouse III—Paul Sabatier, CNRS, CNES, Toulouse, France
  \and
              Space Physics and Astronomy Research Unit and Sodankyl\"a Geophysical Observatory, University of Oulu, Oulu, Finland
  \and
              Department of Physics and Astronomy, University of Turku, 20500 Turku, Finland
  \and
              Institut f\"ur Experimentelle und Angewandte Physik, Christian-Albrechts-Universit\"at zu Kiel, 24118 Kiel, Germany}

\titlerunning{GLE73: First observations}  

 
  \abstract
   {}
   {The first relativistic solar proton event of solar cycle 25 (SC25) was detected on 28 October 2021 by neutron monitors (NMs) on the ground and particle detectors onboard spacecraft in the near-Earth space. This is the first ground level enhancement (GLE) of the current cycle. A detailed reconstruction of the NM response together with the identification of the solar eruption that generated these particles is investigated based on in-situ and remote-sensing measurements.}
   {In-situ proton observations from a few MeV to $\sim$500 MeV were combined with the detection of a solar flare in soft X-rays (SXRs), a coronal mass ejection (CME), radio bursts and extreme ultraviolet (EUV) observations to identify the solar origin of the GLE. Timing analysis was performed and a relation to the solar sources was outlined.}
   {GLE73 reached a maximum particle rigidity of $\sim$2.4 GV and is associated with type III, type II, type IV radio bursts and an EUV wave. A diversity of time profiles recorded by NMs was observed. This points to an anisotropic nature of the event. The peak flux at E$>$10 MeV was only $\sim$ 30 pfu and remained at this level for several days. The release time of $\geq$1 GV particles was found to be $\sim$15:40 UT. GLE73 had a moderately hard rigidity spectrum at very high energies ($\gamma \sim$5.5). Comparison of GLE73 to previous GLEs with similar solar drivers is performed.}
   {}


   \maketitle
%

\section{Introduction}
\label{sec:intro}

Ground Level Enhancements (GLEs) represent the high-energy tail of Solar Energetic Particle (SEP) events. GLEs require acceleration processes capable of producing $\geq$ 1 GV (in rigidity) particles with sufficient intensity to allow their secondary products to reach the terrestrial ground and be detected by neutron monitors (NMs) \citep[e.g.][and references therein]{Poluianov2017}. Due to their fast propagation, relativistic protons in GLEs are particularly useful for the identification of SEP sources at the Sun \citep{aschwanden2012gev}. The relationship between manifestations of solar activity and energetic protons has been investigated in a series of works \citep[e.g.][]{belov2005proton, gopalswamy2012properties,makela2015estimating, firoz2019relation,Kouloumvakos2019}. However, given the relation of GLEs to both strong solar flares and fast and wide CMEs, usually, their acceleration site cannot be unambiguously identified. Detailed studies of specific GLE events have been conducted \citep[e.g.][]{bombardieri2008improved, mishev2018first} but the conditions and processes that lead to such strong SEP events are still not completely understood. GLEs usually have a \textit{gradual} proton component with E$>$10 MeV that lasts for several days and leads to a significant SEP peak flux. Hence GLEs are thought to be dominated by CME-driven shocks \citep[see, e.g.,][]{kahler2012comparison,nitta2012special}. On the other hand, studies of the timing of GLE events have shown evidence for two distinct components, with one being driven by re-connection processes leading to the so-called prompt component (PC) and the other associated with the expanding CME-driven shock that gives ground to the delayed component (DC) \citep[e.g.][]{vashenyuk2006some,mccracken2008investigation,moraal2012time}. Therefore, until today the debate about the exact nature of GLE mechanisms is still ongoing \citep[see e.g.][]{kouloumvakos2020evidence, kocharov2021multiple}.

GLEs are rare (i.e. only 73 events in $\sim$ 80 years of observations)\footnote{\url{https://gle.oulu.fi/}} with a rate of $\sim$ 0.9 events per year \citep{vainio2017solar}. These events have been primarily recorded by NMs on the ground, and their lower energy components were seen by spacecraft in the near-Earth space. Thus, their analysis was hampered by the lack of identification in other vantage points within the heliosphere. However, in recent years, with the launch of the Solar Terrestrial Relations Observatory (STEREO) twin mission \citep{kaiser2008stereo} and the landing of the Mars Science Laboratory (MSL) on Mars \citep{grotzinger2012mars}, GLE71 (17 May 2012) \& GLE72 (10 September 2017) have been identified and investigated as multi-spacecraft events \citep[see e.g.][]{Rouillard2016,battarbee2018multi,guo2018modeling,cohen2018ground}. Adding to this, GLEs have only been investigated based on recordings made in the inner heliosphere for a handful of cases \citep[e.g.][]{cliver2006unusual,reames2013spatial}. Nonetheless, missions of the present day like Solar Orbiter \citep[SolO;][]{muller2020solar}, Parker Solar Probe \citep[PSP;][]{fox2016solar} and BepiColombo \citep{benkhoff2010bepicolombo} may provide a new view on open scientific questions on the origin of relativistic particles since those offer concurrent measurements of protons and complementary electromagnetic observations at a set of vantage points in the inner heliosphere. The present letter combines measurements of GLE73 --the first such event recorded in SC25-- at the near-Earth space and on the ground together with observations of the CME evolution, context solar information and modelling of SEPs based on NM recordings.
\vspace{-3mm}

\section{Observations} \label{sec:ob}

\subsection{Overview} \label{subsec:overview}

\begin{figure}[h!]
\centering
\includegraphics[width=0.85\columnwidth]{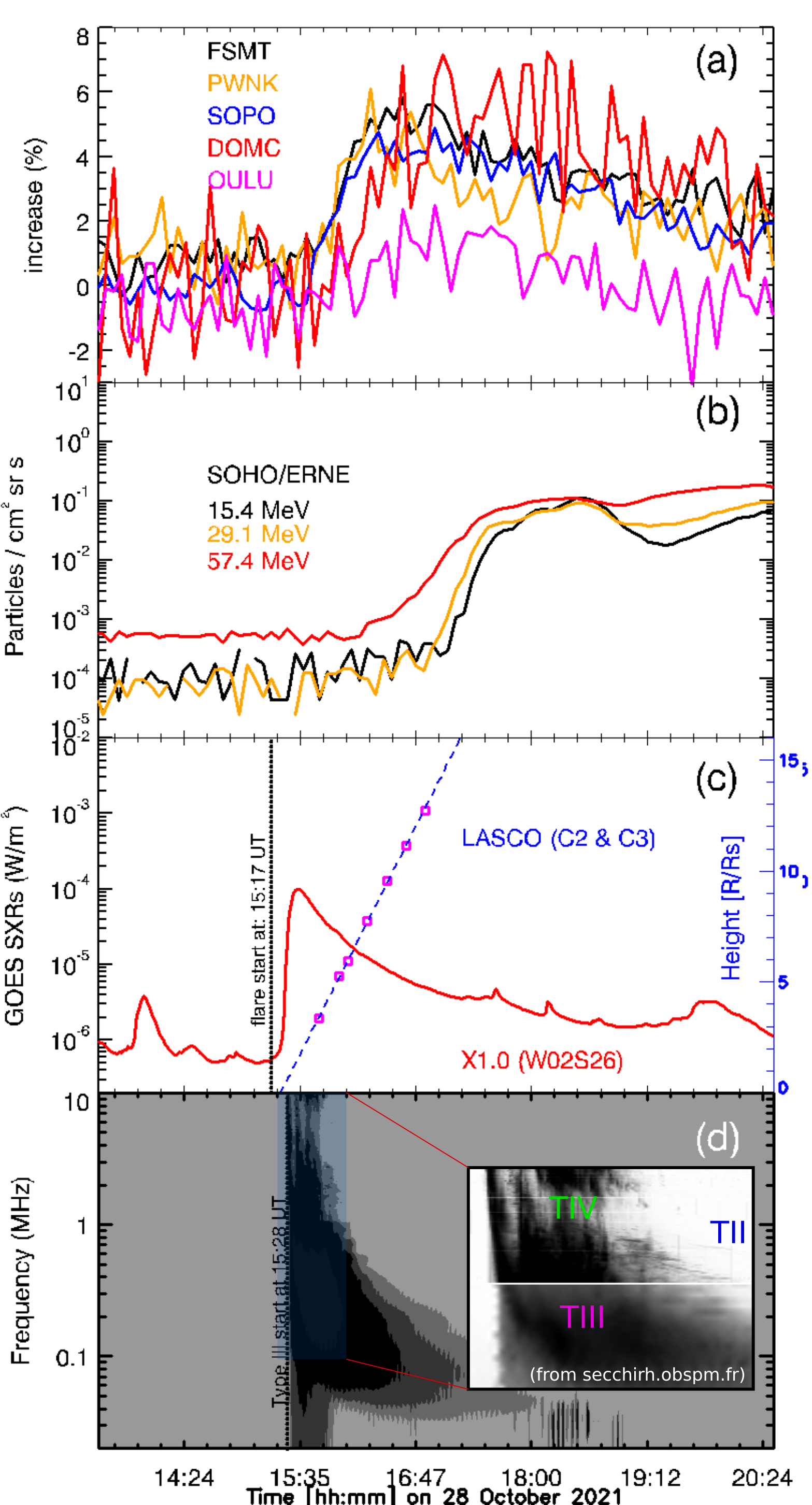}
\caption{GLE73 on 28 October 2021. From (a) to (d): The increase (\%) of several NMs based on 5-min de-trended NM data; the SOHO/ERNE proton flux. The SXRs flux observed by GOES, denoting an X1.0 solar flare (red curve; left axis).The black dashed vertical line corresponds to the start time of the flare. The height of the WL shock's is shown with the magenta squares from measurements at the plane-of-sky near the CME leading-edge. The dashed blue line is a linear fit to the height and extrapolated back to the surface of the Sun. The dynamic radio spectrum was observed by Wind/WAVES. The dashed black line corresponds to the start time of the identified type III burst at $\sim$15:28~UT. The inset figure denotes the type II burst (in blue), the storm of type III bursts (in magenta) and the type IV burst (in green).} 
\label{fig:fig1}
\end{figure}

The first GLE event (GLE73) of SC25 was observed by several neutron monitors around the Earth (see Table \ref{tab:table1}), on 28 October 2021. Figure~\ref{fig:fig1} shows an overview of observations during the GLE73 event. The peak intensity was maximum for the two conventional NM stations located on the Antartic  plateau, $\sim$7.3\% for DOMC (Dome C NM at Concordia station) and 5.4\% for SOPO (South Pole). Bare (lead-free) NMs at the same sites detected a higher response (14.0\% for DOMB and 6.6\% for SOPB). Energetic protons were also observed by the Solar and Heliospheric Observatory (SOHO)/Energetic and Relativistic Nuclei and Electron (ERNE) \citep{torsti1995energetic} at a range of energies (see Appendix \ref{appendix:A}). Figure \ref{fig:fig1} (b) depicts three proton channels of ERNE. GLE73 was associated with an X1.0 class flare starting at 15:17~UT and peaking at 15:35~UT (see Figure~\ref{fig:fig1} (c)). The source active region NOAA AR12887 was located at W02S26 (in Heliographic Stonyhurst (HGS) coordinate system at 15:20~UT) as observed by Earth's viewpoint. In addition, from metric to kilo-metric wavelengths (radio domain) type III, type II and IV radio bursts were also observed in association with the solar event. In particular, the start time of the first type III is marked at 15:28~UT (see Figure \ref{fig:fig1}(d)) which further coincides with the start of a type II radio burst\footnote{\url{http://soleil.i4ds.ch/solarradio/data/BurstLists/2010-yyyy_Monstein/2021/e-CALLISTO_2021_10.txt}}. The group of type III bursts is evident from $\sim$15:30-15:50~UT, whereas a metric type IV radio burst is also marked at $\sim$15:37~UT (see the inset in Figure \ref{fig:fig1}(d)). 

\begin{figure*}[h!]
\centering
\includegraphics[scale=0.8]{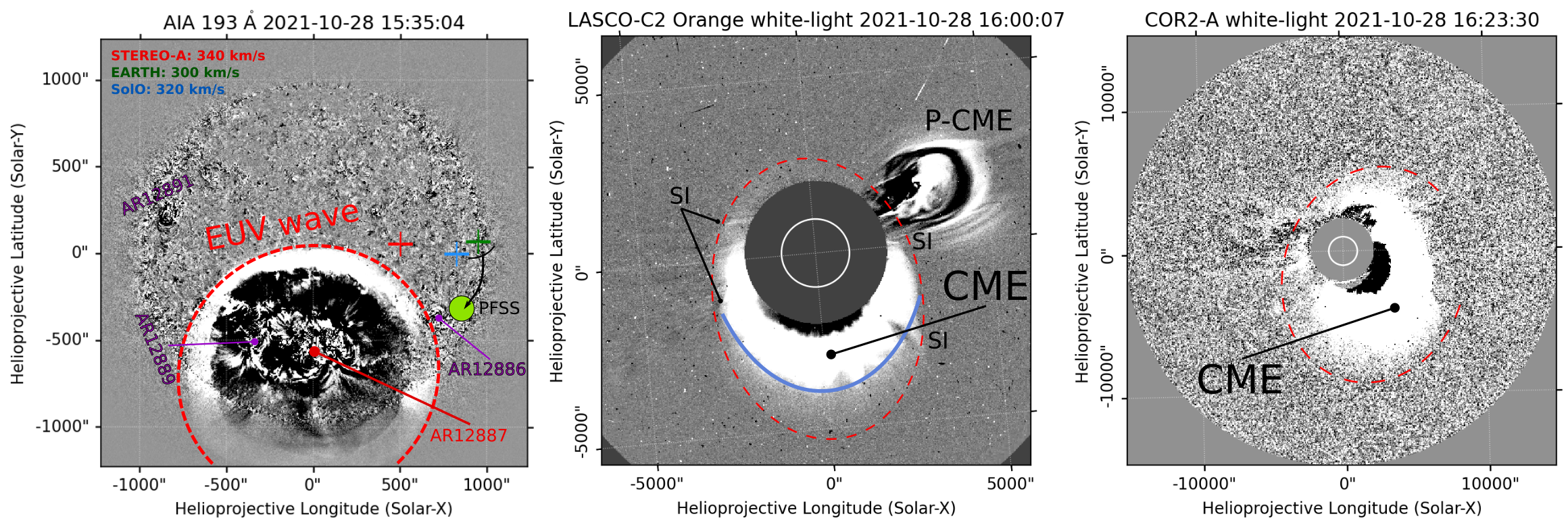}
\caption{Selected snapshots of EUV and WL coronagraphic observations before and during the GLE73 event on 28 October 2021. The left panel shows a running-difference image in EUV from SDO/AIA at 193~\AA. The EUV wave is encircled to indicate its location. The footpoints of the Parker spirals connected to PSP (purple), STEREO-A (red), SolO (blue), and Earth (green) are shown with the coloured crosses. Most of the footpoints of the magnetic field lines connected to Earth (from PFSS) gather close to the region highlighted with a green circle. We indicate the location of ARs that the pressure/shock wave interacted during its expansion. The middle and right panels show running-difference images from LASCO-C2 and STEREO-A/COR2 coronagraphs, respectively. We outline the location of the shock wave and indicate the CME. We indicate the locations that the pressure/shock wave interacted with coronal streamers (SI) and the previous CME (P-CME).
} 
\label{fig:EUV_WL}
\end{figure*}


\subsection{The CME and the EUV wave} \label{sec:rs}

GLE73 was also associated with an Extreme Ultraviolet (EUV) wave that was observed in the low corona by EUV imagers (i.e. the Atmospheric Imaging Assembly (AIA) of Solar Dynamics Observatory \citep[SDO;][]{lemen2011atmospheric} and  \cite[STEREO-A/EUVI;][]{howard2008sun}), and a CME and white light (WL) shock wave that was observed higher in the corona by the SOHO/LASCO \citep{brueckner1995large} and STEREO-A coronagraphs \citep{howard2008sun}. In Figure~\ref{fig:EUV_WL}, we show remote-sensing observations during the solar event. EUV observations show a classic picture of an EUV wave, namely a circular propagating bright front, to form at $\sim$15:28~UT. The EUV wave expanded coherently toward every direction and evolved clearly as a global wave from the Earth's viewpoint, engulfing the visible disk by 16:20~UT (see the complementary movie).

The CME was well observed by two different spacecraft, namely STEREO-A and SOHO, that were separated by 38$^\circ$ (see also Figure \ref{fig:sc_pos}). At LASCO/C2 and STEREO-A/COR1 the CME was observed for the first time at 15:48 UT (at $\sim$2.83~$R_\sun$ and a Position Angle (PA): $\sim$185$^\circ$) and at 15:36 UT (at $\sim$1.91~$R_\sun$ and PA:$\sim$230$^\circ$), respectively. Both viewpoints reveal the emergence of a broad CME forming a halo and a clear pressure wave in front moving faster than the erupting plasma. The wave appears to interact with coronal streamers located on the CME flanks (see Figure~\ref{fig:EUV_WL}). There is also a narrow and slow CME that erupted a few hours before GLE73, from an AR located just behind the west solar limb. The west flank of the WL pressure/shock seems to interact with the southern section of this previous CME's legs. From a linear fit to the height-time measurements, we obtained a plane-of-sky CME speed at its leading-edge (PA:$\sim$185$^\circ$) of the combined LASCO/C2 \& C3 field of view of $\sim$1240$\pm$40~km/s (see Figure \ref{fig:fig1}(c)). At the same direction we obtained for the WL shock a plane-of-sky speed of $\sim$1640$\pm$40~km/s. 

\begin{figure}[h!]
\centering
\includegraphics[width=0.85\columnwidth]{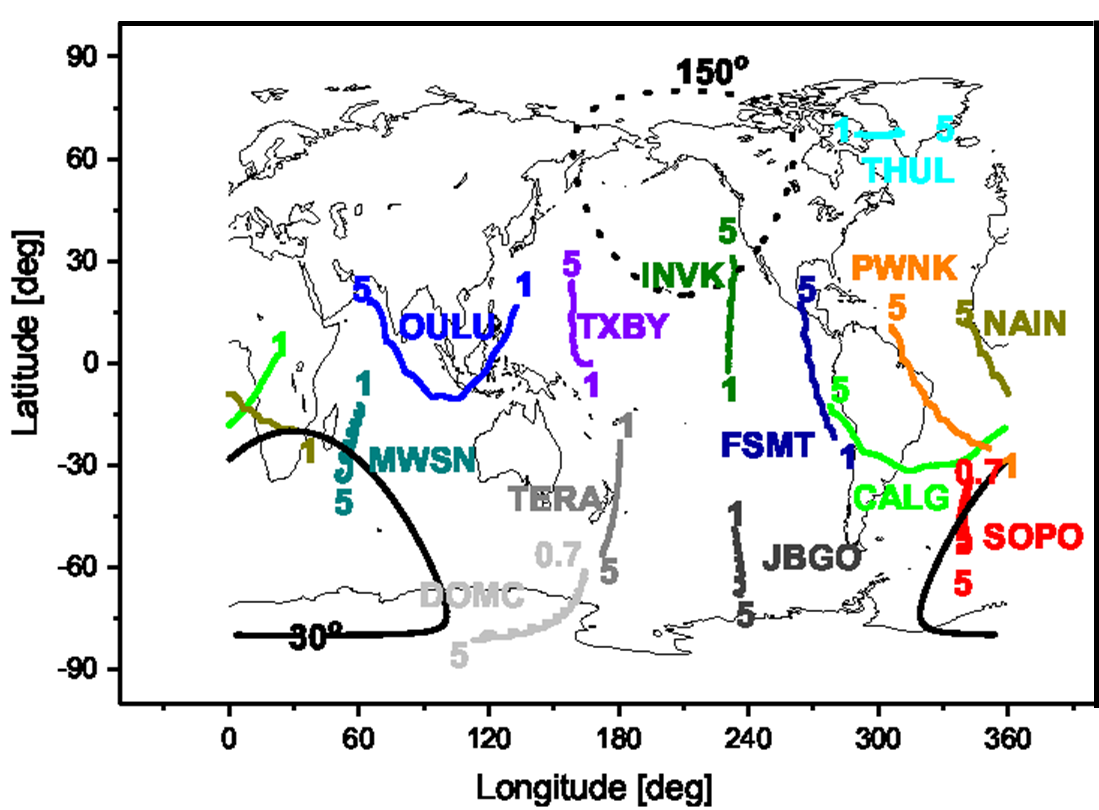}
\caption{Viewing directions of neutron monitors in the GSE coordinates at around the onset of GLE73 at 15:50 UT on 28 October 2021. Geomagnetic conditions were slightly disturbed (Kp = 1.0). The color lines, acronyms and numbers depict the asymptotic directions, NM stations and rigidities at a range 1-5 GV, respectively. The line of equal pitch angles relative to the anisotropy axis is plotted for 30$^\circ$ for the sunward direction (solid black line) and for 150$^\circ$ for the anti-sunward direction (dashed black line), respectively.} 
\label{fig:fig2}
\end{figure}
\vspace{-4mm}

\subsection{Neutron Monitor Data}
\label{sec:neutron}

During GLE73  differences in the time profiles of the cosmic-ray intensity are evident, as revealed by the Fort Smith (FSMT), Dome C Concordia (DOMC), South Pole (SOPO), Oulu (OULU) and Peawanuck (PWNK) NMs presented in Figure \ref{fig:fig1} (a). Herein, we use five-minute integrated de-trended NM data retrieved from the international GLE database\footnote{\url{https://gle.oulu.fi/}}. NM data are also presented in the neutron monitor database (NMDB). One can see that the event revealed a typical gradual increase and moderate anisotropy (see details in Sections \ref{sec:NMres} \& \ref{sec:NMcom}) during the onset since a moderate count-rate increase is recorded by stations looking in the sunward direction (FSMT, PWNK, SOPO). As can be seen in Figure \ref{fig:fig1}, during GLE73 the flux remained above the background level for almost 4.5~hours. The NM situated at high-altitude polar stations, i.e. DOMC and SOPO recorded the greatest count rate increases. The rapid rise as shown by the FSMT, SOPO and PWNK NMs intensity time-profile (Figure \ref{fig:fig1}) indicates that energetic protons had reasonable access to the Sun-Earth-connecting field lines. For twelve NMs and the two bare NMs the onset and peak time, as well as the maximum increase (in \%) were calculated using the de-trended NM data \citep{usoskin2020revised}, as discussed in Appendix \ref{appendix:B}. All results are presented in Table \ref{tab:table1}.

\vspace{-3mm}
\section{Results}

\subsection{Modeling the neutron monitors response}
\label{sec:NMres}
The analysis of GLEs based on NM data consists of several consecutive steps \citep[see][]{smart2000magnetospheric}. The detailed description of the model used in this work is given in \cite{mishev2014analysis} and \cite{Mishev2021AP}. A method that has been recently applied to a series of GLEs \citep[i.e.][]{mishev2017assessment,mishev2018first}. Figure \ref{fig:fig2} shows the calculated viewing directions of the NMs used in this analysis at around the onset of GLE73 (15:50 UT) for particles of 1 to 5 GV, accordingly 0.7-- 5 GV for the high-altitude polar NMs, whilst in the analysis the rigidity range up to 20 GV was implied. The FSMT, SOPO, PWNK and Nain (NAIN) NMs possess viewing directions that are close to the nominal sunward direction, whilst Inuvik (INVK) NM had viewing direction close to the nominal anti-sunward direction. The SOPO and FSMT NMs observed an earlier onset, with a more rapid rise being exhibited by FSMT and PWNK, while INVK revealed gradual rise. Naturally, it is related to the location of those station(s). 

\begin{figure}[h!]
\centering
\includegraphics[width=\columnwidth]{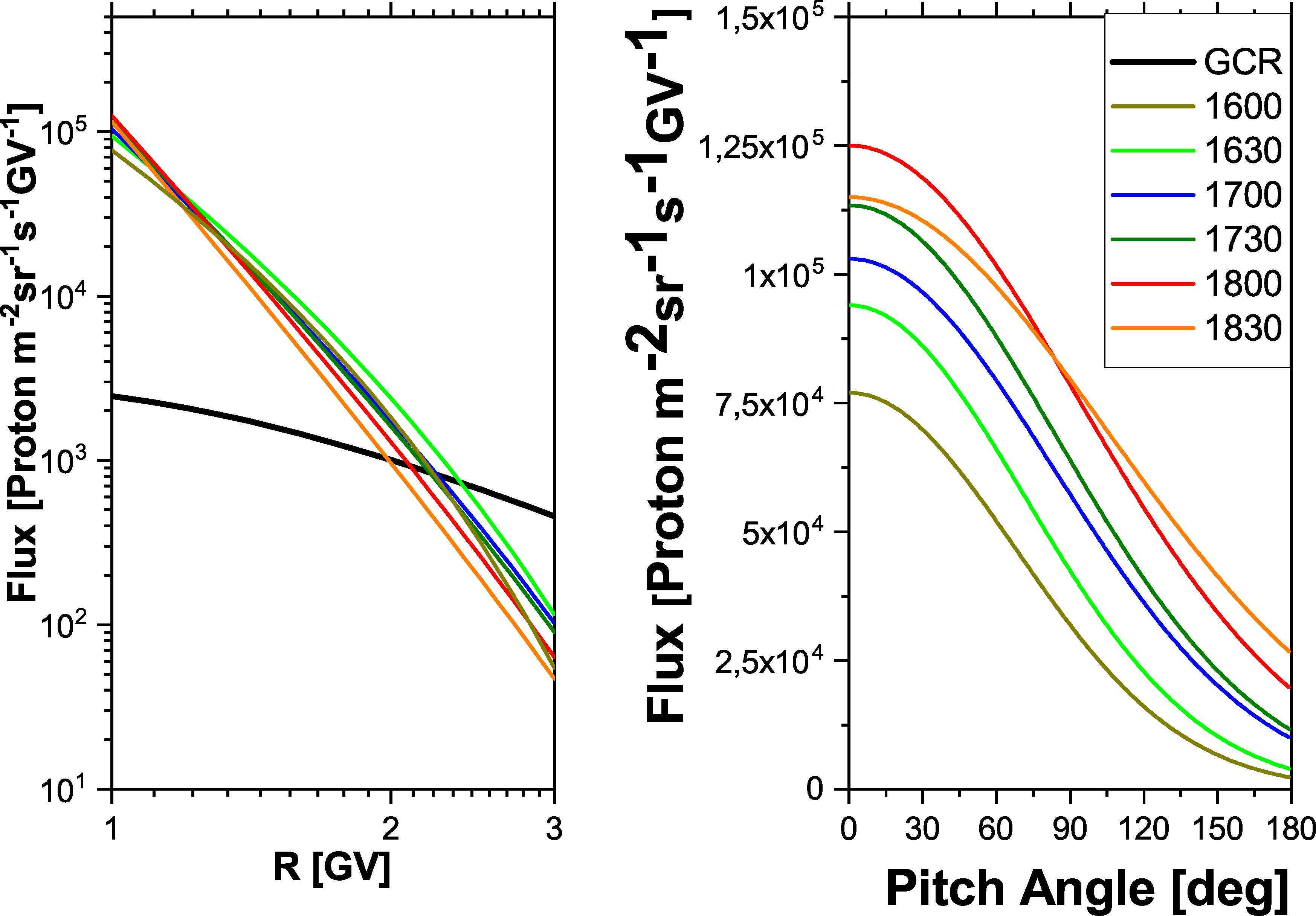}
\caption{Derived SEP rigidity spectra (left panel) and PADs (right panel)
during GLE73 on 28 October 2021. The solid black line denotes the
galactic cosmic ray flux, which corresponds to the time period of the GLE 73 occurrence (see text for details).
All time in the legend are in UT and refer to the start of the corresponding
five minute interval over which the data are integrated.} 
\label{fig:spectra}
\end{figure}

Employing the model presented in Appendix \ref{appendix:B}, we derived the spectra (see Eq.~(\ref{eq:eq1})), pitch-angle distribution (PAD) and apparent source (see Eq.~(\ref{eq:eq2})) position of the solar protons during the main phase of GLE73. The spectra gradually softened in the course of the event, specifically during the initial and  main phases of the event, the latter corresponding to about 17:30--18:20 UT, that is during the peak intensity of the event \citep[e.g. see the discussion in][]{Mishev2021}. The results are presented in Figure \ref{fig:spectra} and the details are given in Table \ref{table:spectra}. The derived spectra are moderately hard with moderate steepening ($\delta\gamma$). Moreover, we derived moderately anisotropic angular distribution fitted with a function similar to a Gaussian, without
any signature of protons arriving from the anti-sunward direction,
nor a complicated PAD as depicted in \citet{mishev2014analysis}. The derived angular distribution gradually broadens during the main phase of the event. 

\vspace{-5mm}
\subsection{In-situ particles}
GLE73 was clearly recorded by particle instruments on near-Earth orbiting spacecraft such as ERNE onboard SOHO, the Space Environment in Situ Suite (SEISS) on Geostationary Operational Environmental Satellite (GOES) \citep{kress2020goes} and the High Energy Telescope (HET) of the Energetic Particle Detector (EPD) on Solar Orbiter (SolO) \citep{rodriguez2020energetic}. Figure \ref{fig:fig1}(b) shows the recordings from SOHO/ERNE for a set of three energy channels with an effective energy of 15.4, 29.1 \& 57.4 MeV. Figure \ref{fig:GOEs} in Appendix \ref{appendix:A} shows the 5-min averaged recordings of solar particles on GOES/SEISS [6.5-500 MeV]; SOHO/ERNE [15.4-57.4 MeV] and SolO/HET [13.68-89.46 MeV] together with the recordings of the BCB-counter of SolO/HET [E$>$157 MeV; \cite[see details in][]{freiherr2021radial}]. In addition, Figure \ref{fig:sc_pos} shows the relative position of spacecraft at 15:15~UT and indicates that in addition to GOES and SOHO, SolO was also close to Earth at a distance of 0.80~AU (astronomical units).

\subsection{Relation to solar sources}

For the first arriving particles it is possible to perform time-shifting analysis \citep[TSA;][]{vainio2013first} to infer their release time at the Sun (Solar Release Time; SRT). A low-end energy limit of particles recorded by a sea level NM station is $\sim$1 GV (i.e. 433 MeV) thus the corresponding mean velocity for such energetic protons would be $u = 0.73c$. For GLE73 particles with rigidities up to $\sim$2.4 GV (1.6 GeV) have been identified, with a mean velocity of $u = 0.93c$. The length of the Parker spiral $L$ can be computed based on the solar wind speed during the event. During GLE73, the solar wind speed was slow $V_{SW}$=300~km/s, leading to $L$ = 1.28 AU. For the first arriving particles we assume scatter-free propagation and calculated the expected SRT of the relativistic protons, $t_{rel}$, adding 500s for comparison with remote-sensing measurements at 1 AU (e.g. radio observations) \citep{papaioannou2014first}. For SOPO NM station, that registered the earlier onset, we obtained $t_{onset}$ = 15:45 UT (see Table \ref{tab:table1}). The travel time of the relativistic protons of $\sim$2.4~GV was calculated to be $\sim$11 min and the corresponding anticipated  $t_{rel}$ $\sim$15:42 UT. For a set of rigidities $\geq$1 GV $t_{rel}$ ranges from 15:39-15:42~UT. Since 5-min resolution NM data are used, there is a 5-min uncertainty in these calculations.

From SDO/AIA images we track the expansion of the EUV wave toward the footpoints of the magnetic field lines connected to Earth that we determined using the Potential Field Source Surface (PFSS) model and global photospheric magnetic maps (see Appendix~\ref{appendix:PFSS}). We find that the footpoints were located $\sim$72$^\circ$ west from AR12887. The release time of the relativistic particles for the GLE73 seems to connect well to the time that the EUV wave passed by the location of the footpoints magnetically connected to Earth (see Figure~\ref{fig:aia} at $\sim$15:39 UT). Comparing with the SXR and radio observations, we find that the release of $\sim$2.4 GV particles ($\sim$15:42~UT) is $\sim$5~minutes after the flare peak time and 12~minutes after the start of the first type III and the type II ratio burst (Figure~\ref{fig:fig1}). Around the release time of energetic protons (R$\geq$ 1GV which ranges between $\sim$15:39-15:42~UT) there is radio emission from a group of the type IIIs and a moving type VI radio bursts (see Figure~\ref{fig:fig1}). At the release time of the $\sim$2.4 GV particles the WL shock is located at a height of $\sim$2.32~$R_\sun$. Table \ref{tab:timeline} provides a timeline of events during GLE73 based on the measurements and calculations.

\subsection{Comparison with other GLEs}
\label{sec:NMcom}
\begin{figure}[h!]
\centering
\includegraphics[width=0.90\columnwidth]{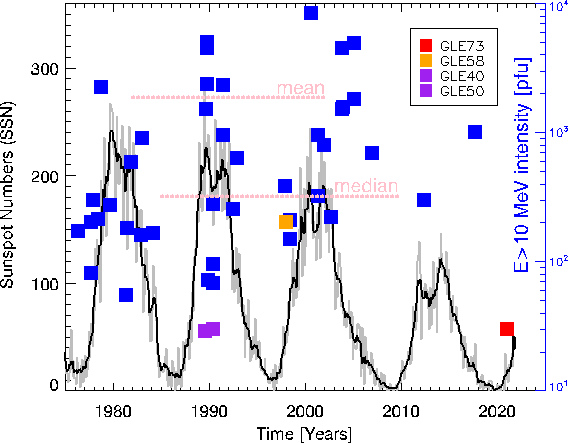}
\caption{Peak proton flux ($I_{P}$) at E$>$10 MeV for GLEs occurring in 1976–2021 (blue squares). GLE58 (orange square), GLE40 \& GLE50 (purple squares) and GLE73 (red square) are indicated (see text). Mean (1850 pfu) and median (321 pfu) values of the peak proton flux are presented as pink dashed horizontal lines. Monthly and smoothed monthly sunspot numbers are shown as grey and black lines.} 
\label{fig:sc}
\end{figure}

There are only 5 GLEs since 1976 that were associated with an $\leq$X1.0 SXR flare (i.e. GLE30, GLE32, GLE58, GLE62 \& GLE71). However, only GLE58 is associated with a central (E09) X1.0 flare. Figure \ref{fig:sc} shows the time distribution of all GLEs since 1976 with respect to E$>$10 MeV proton peak flux $I_{P}$ detected by the series of GOES satellites. Despite the similar flare flux and position, GLE58 (orange square) has an $I_{P}$ 6.7 times larger than GLE73 (red square). GLE40 \& GLE50 (purple squares) have similar $I_{P}$ ($\sim$30 pfu) but both were limb events ($>$W85). 
Around the time of release of the 2.4 GV particles the height of the CME and the WL shock was $\sim$1.84~$R_\sun$ and $\sim$2.32~$R_\sun$, respectively. Both values are lower than the mean values reported for other poorly connected GLEs \citep[see][]{gopalswamy2012properties}. Also the median plane-of-sky (projected) speed for GLEs is $\sim$1810 km/s \citep[see][]{gopalswamy2012properties} whereas the GLE73 CME speed from LASCO C2 \& C3 was estimated to be $\sim$1240 km/s and the plane-of-sky speed of the WL shock was $\sim$1640 km/s.

\vspace{-3mm}

\section{Conclusions}

In this work a summary of observations for GLE73 that took place on 28 October 2021 --the first such event of SC25-- is presented. Detailed modeling and reconstruction of the spectral and angular characteristics of high-energy SEPs in the vicinity of the Earth was performed. Ground-based NMs, together with space-borne data were employed in the corresponding data analysis. One of the characteristic aspects of this GLE is its association with a central-disk (W02) X1.0 flare (fairly untypical for GLEs) and a CME (of $\sim$1240 km/s) driving a WL shock (of $\sim$1640 km/s). The main results of the study are:

\begin{enumerate}
    \item During the main phase of GLE73 the rigidity spectrum was moderately hard ($\gamma \sim$5.5), with significant steepness $\delta\gamma \sim$0.4. During this stage of the event the derived PAD was relatively wide ($\sigma^{2}\approx$4.5 rad$^{2}$).
    
    \item  The event was characterized by a directional particle flux arriving from the sunward direction, hence GLE73 was characterized by a relatively strong anisotropy. 
    
    \item The SRT of the very high energy particles was found to be $\sim$15:40~UT and around this SRT the CME-driven shock was located at a height of $\sim$2.32 ($\pm0.2$)~$R_\sun$.
    
    \item Timing of the EUV wave evolution towards the field lines magnetically connected to Earth and the inferred release time of high energy protons seem to be in good agreement.
\end{enumerate}


\begin{acknowledgements}
The authors acknowledge the International Space Science Institute and the supported International Team 441: \textit{HEROIC}.

AP acknowledge support from NASA/LWS project NNH19ZDA001N-LWS. 

AK, RV and JG acknowledges financial support from the European Union’s Horizon 2020 research and innovation programme under grant agreement No.\ 101004159 (SERPENTINE). AK and AR acknowledge the ANR COROSHOCK project (ANR-17-CE31-0006-01).

AM and AP acknowledge the support by the Academy of Finland (project 330064 QUASARE). 

KH acknowledges the support of the DFG priority program SPP 1992 “Exploring the Diversity of Extrasolar Planets (HE 8392/1-1)”. AP and KH also acknowledge the supported International Team 464: \textit{ETERNAL}.

IU acknowledges a partial support from the Academy of Finland (project ESPERA No. 321882).

RV and JG acknowledge the support of Academy of Finland (FORESAIL, grants 312357 and 336809).

We acknowledge the NMDB database (\url{www.nmdb.eu}) founded under the European Union's FP7 programme (contract no. 213007), and the PIs of individual neutron monitors. Italian polar program PNRA (via the LTCPAA PNRA 2015/AC3 and the BSRN PNRA OSS-06 projects), the French Polar Institute IPEV and FINNARP are acknowledged for the hosting of DOMB/DOMC NMs.

\end{acknowledgements}

\vspace{-7mm}

\bibliographystyle{aa}
\bibliography{references}

\begin{appendix} 

\section{Near Earth measurements of the SEP event of 28 October 2021} \label{appendix:A}

\begin{figure}[h!]
\centering
\includegraphics[width=0.87\columnwidth]{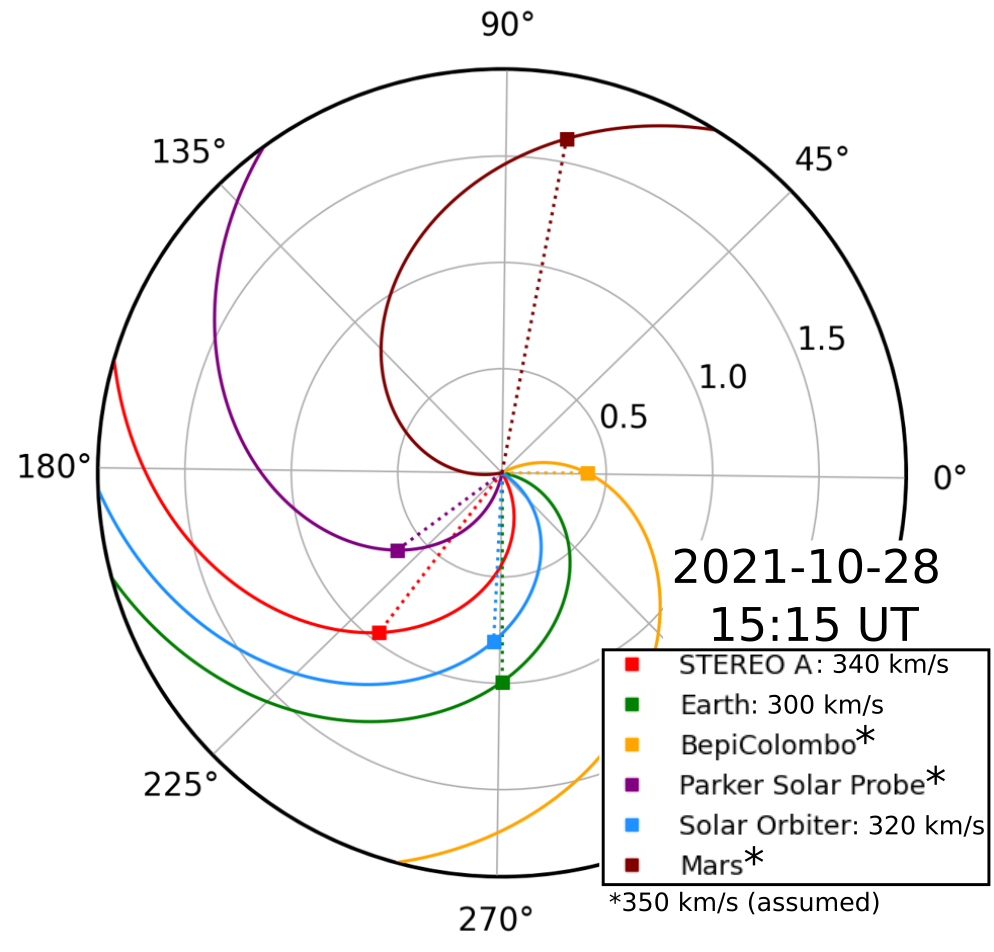}
\caption{A view of the ecliptic plane from solar north showing the positions of various spacecraft on 28 October 2021 at 15:15~UT. The Parker spirals are shown for each spacecraft. From the Solar MAgnetic Connection Haus tool (https://solar-mach.github.io/).} 
\label{fig:sc_pos}
\end{figure}

At the time of the GLE73, SolO, STEREO-A, and PSP were trailing Earth by -3$^\circ$, -38$^\circ$ and -54$^\circ$, respectively, while BepiColombo was leading Earth by 90$^\circ$. Figure~\ref{fig:sc_pos} shows the positions of various spacecraft at the heliosphere and the Parker spirals connecting at each location. A measured solar wind speed was used for each spacecraft when available\footnote{For SolO we used the solar wind speed during October 30 before the shock arrival.}, or else a speed of 350~km/s was assumed. Using the measured solar wind speeds shown at the legend of Figure~\ref{fig:sc_pos}, for Earth, STEREO-A, and SolO, we calculated the location of the footpoints of the nominal Parker spirals. The footpoints connected to Earth, STEREO-A, and SolO were located  at W81N05, W31N07, and W63N02, respectively (in HGS system at 15:20~UT).

\begin{figure}[h!]
\centering
\includegraphics[width=0.88\columnwidth]{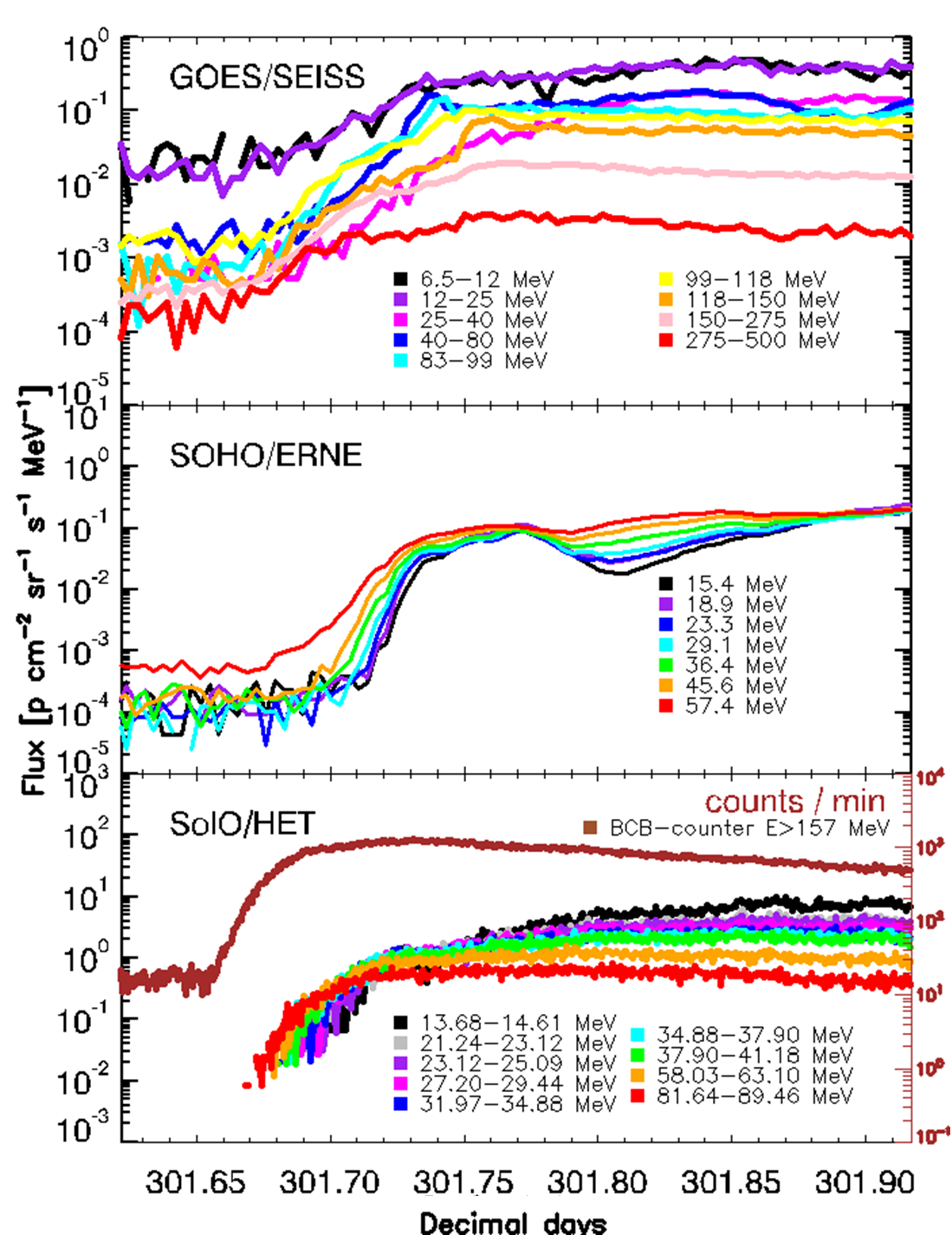}
\caption{Energetic particle recordings of GLE73 in the near Earth space, (from top to bottom) 5-min averaged GOES/SEISS differential fluxes; SOHO/ERNE fluxes and SolO/HET measurements including the recordings of the SolO/HET/BCB-counter.} 
\label{fig:GOEs}
\end{figure}

GLE73 was clearly recorded by near-Earth spacecraft  GOES and SOHO, as well as, SolO that was in a favorable position to record it. The analysis of GLE73 using all heliospheric vantage points is beyond the scope of this letter, and we concentrate instead on the near-Earth spacecraft and SolO, which is the least separated from Earth (see Figure~\ref{fig:sc_pos}). The time history of SEP measurements during the GLE73 as recorded (from top to bottom) from GOES/SEISS [6.5-500 MeV], SOHO/ERNE [15.4-57.4 MeV] and SolO/HET [13.68-89.46 MeV \& BCB-counter [E>157 MeV; counts/min]  \citep{freiherr2021radial}] are presented in Figure \ref{fig:GOEs}.  High energy protons at each spacecraft (all indicated with a red line in each panel) seem to have a prompt increase, with GOES/P10 (275-500 MeV) having an onset time at 15:55~UT, SOHO/ERNE (at 57.4 MeV) records the event at 16:18~UT and the BCB-counter of SolO (Figure \ref{fig:GOEs}, third panel from the top; brown line) has an onset time at 15:40~UT. Note, however, that at the lowest energies, there seems to be some high-energy contamination in the GOES channels (see Figure \ref{fig:GOEs}, top panel). 

\section{Magnetic connectivity using the PFSS model} \label{appendix:PFSS}

\begin{figure}[h!]
\centering
\includegraphics[width=\columnwidth]{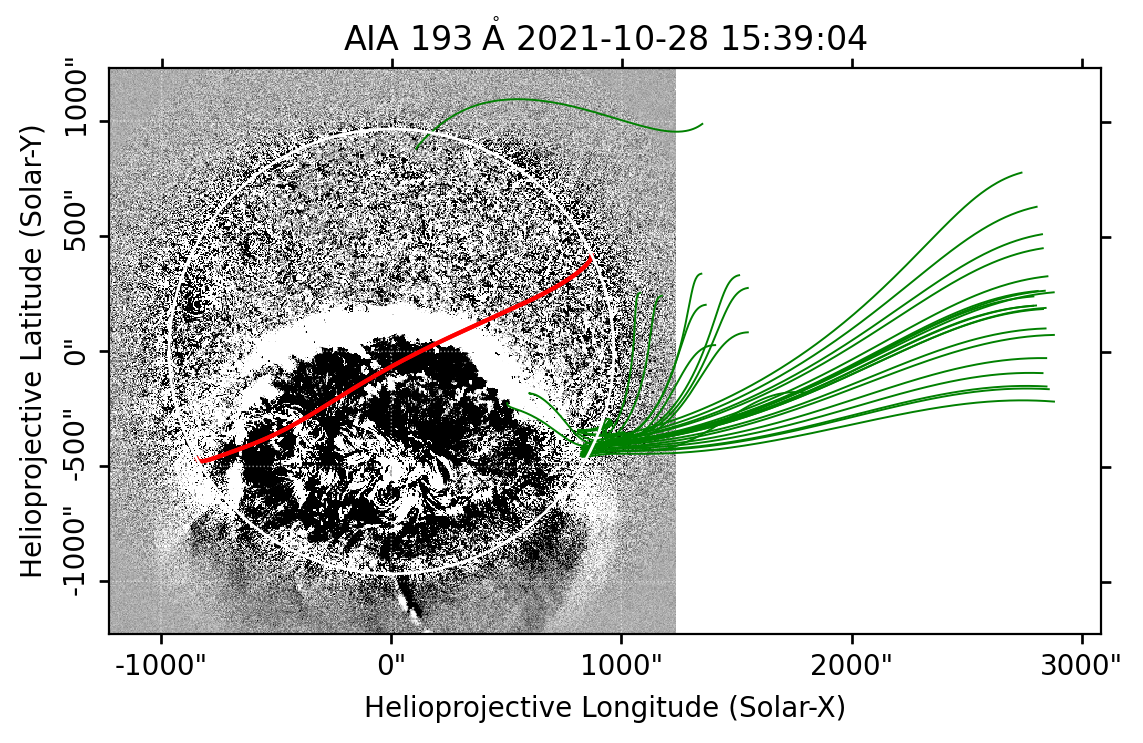}
\caption{Evolution of the EUV wave and PFSS magnetic field lines for Earth (green) are presented at $\sim$15:39 UT. The location of the heliospheric current sheet is shown with the red line. } 
\label{fig:aia}
\end{figure}

Since the magnetic configurations in the low corona are more complex than the simple Parker spiral model employed in Figure~\ref{fig:EUV_WL}, we also used the PFSS model and global photospheric magnetic maps to calculate the magnetic field configuration in the low corona\footnote{\url{http://connect-tool.irap.omp.eu/}}\citep{Rouillard2020}. This gives some further context to the magnetic connectivity of Earth. For the input magnetic maps we used the maps provided by the Air Force Data Assimilative Photospheric Flux Transport (ADAPT) model \citep[][]{Arge2013}. The ADAPT maps are global magnetograms of the photospheric magnetic flux. Then we used the PFSS model and the global maps of the radial magnetic field at the photosphere and calculated the magnetic field from the solar surface to the 3.0~R$_\sun$ which is the assumed height of the source surface. From the location of the footpoint of the Parker spiral at the source surface we determined the field lines connected to Earth. We found that most of the footpoints of the magnetic field lines connected to Earth gather to the west from the AR12886 which was located at W59S19, (e.g., $\sim$57$^\circ$ west from AR12887). Specifically, we found that the average location of the footpoints was at W74S25 and they sparsed for about 10$^\circ$ from this location. In addition, Figure \ref{fig:aia} provides the combined outputs of PFSS and the evolution of the EUV wave at the inferred release time (i.e. 15:39~UT) of the high energy particles ($\geq$1 GV).  

\begin{table}[!ht]
    \centering
\caption[]{Timeline of events for GLE73. }
\label{tab:timeline}

    \begin{tabular}{l c}
    
    \hline
        Event & Time [UT] \\ 
        \hline
        \hline
        SXR onset & 15:17 (1min)\\ 
        Type II onset & 15:28 (1sec) \\ 
        Type III onset (first of the group) & 15:28 (1sec) \\
        EUV wave formation & 15:28 (1min) \\
        SXR peak & 15:35 (1min)\\
        CME first observation in STEREO-A/COR1 & 15:36 ($\sim$5min) \\
        Type IV (m) &  15:37 (1min) \\ 
        SPR Time ($\geq$1 GV) & 15:39 (5min)\\
        EUV wave connection to Earth & 15:40 (1min) \\
        SolO/BCB onset (at E$>$157 MeV) & 15:40 (5min) \\
        GLE onset at South Pole & 15:45 (5min)\\ 
        CME first observation in LASCO/C2 & 15:48 ($\sim$12min) \\
        GOES/P10 onset (at 275-500 MeV) & 15:55 (5min) \\
        SOHO/ERNE onset (at 57.4 MeV) & 16:18 (5min) \\
        \hline
        \multicolumn{2}{p{\columnwidth}}{\textbf{Notes}. All times are Earth times and propagation times for electromagnetic emissions have been considered in this table as explained in the text. The numbers in parenthesis denote the time resolution of the measurements used.}
    \end{tabular}
\end{table}


\section{Analysis of Neutron Monitor measurements} \label{appendix:B}

\begin{figure}[h!]
\centering
\includegraphics[width=\columnwidth]{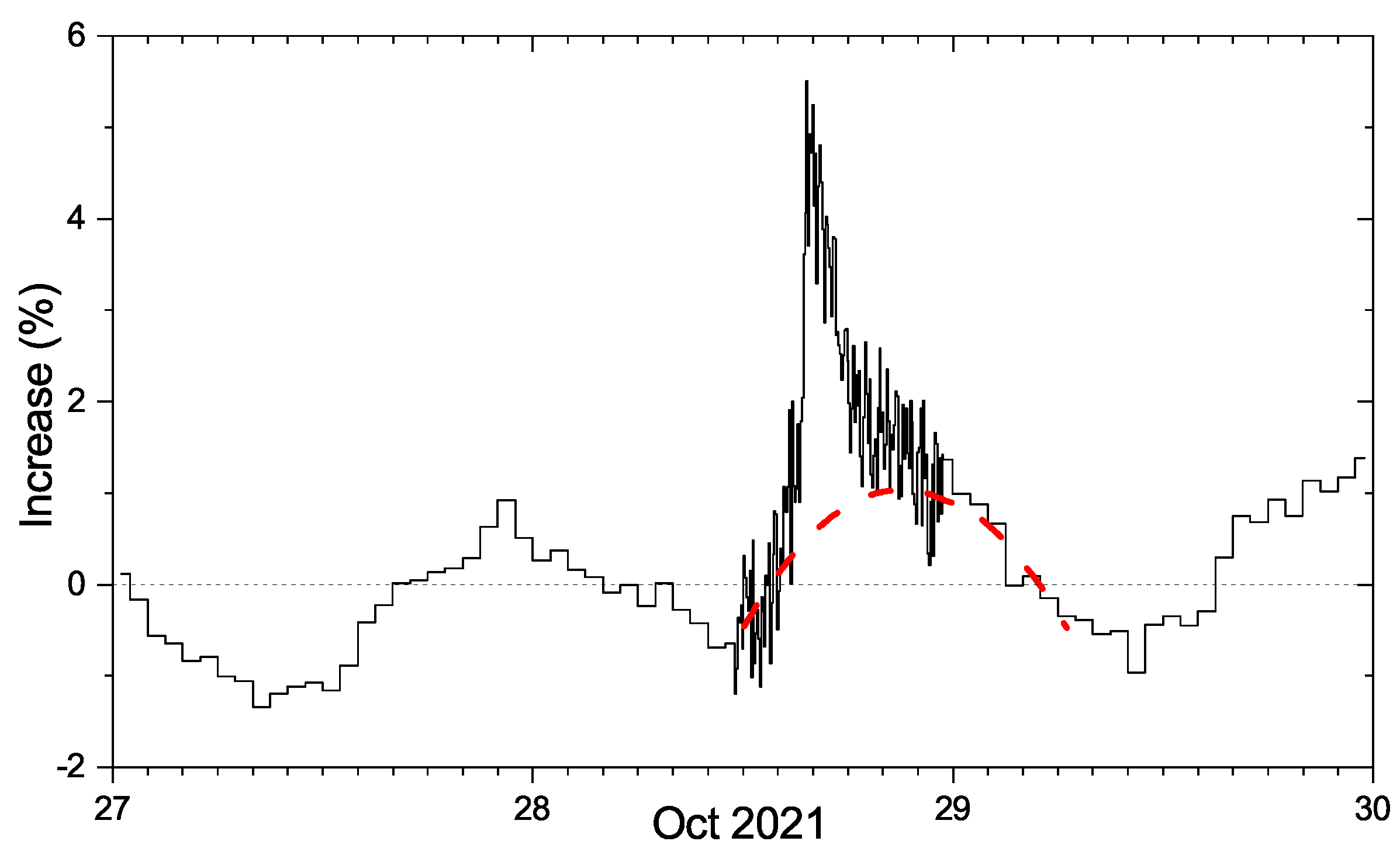}
\caption{Pressure-corrected count rate (5-min data between 12--24 UT of 28-Oct-2021, and hourly otherwise) of Calgary NM, in percent with respect to the pre-increase level of 353 cnts/sec (13--15 UT of 28-Oct-2021) as indicated by the black dotted line. The red dashed line depicts a parabolic-shaped GCR background due to the diurnal wave, with respect to which the GLE strength is calculated. The data is available at IGLED.} 
\label{fig:calg}
\end{figure}

\underline{\textit{\textbf{Measurements}}}
Inspection of the NM data from various stations around the world indicated the presence of particles with a rigidity up to $\sim$2 GV. Newark NM, with a vertical cut-off rigidity of 2.4 GV, recorded an increase of marginal significance that may or may not be related to GLE73. The de-trended NM data \citep{usoskin2020revised} were used in the study. Essentially, the de-trended data account for smooth temporal variability in the baseline, allowing for a clear estimation of the contribution of solar particles in GLEs, free from the effect of short-time variability of galactic cosmic rays (GCRs) due to interplanetary transients and local anisotropy. Figure \ref{fig:calg} illustrates the recordings of Calgary (GALG) NM and shows that GLE73 in the recordings of this station occurred on the background of a strong diurnal wave caused by the local GCR anisotropy.

\begin{figure}[h!]
\centering
\includegraphics[width=\columnwidth]{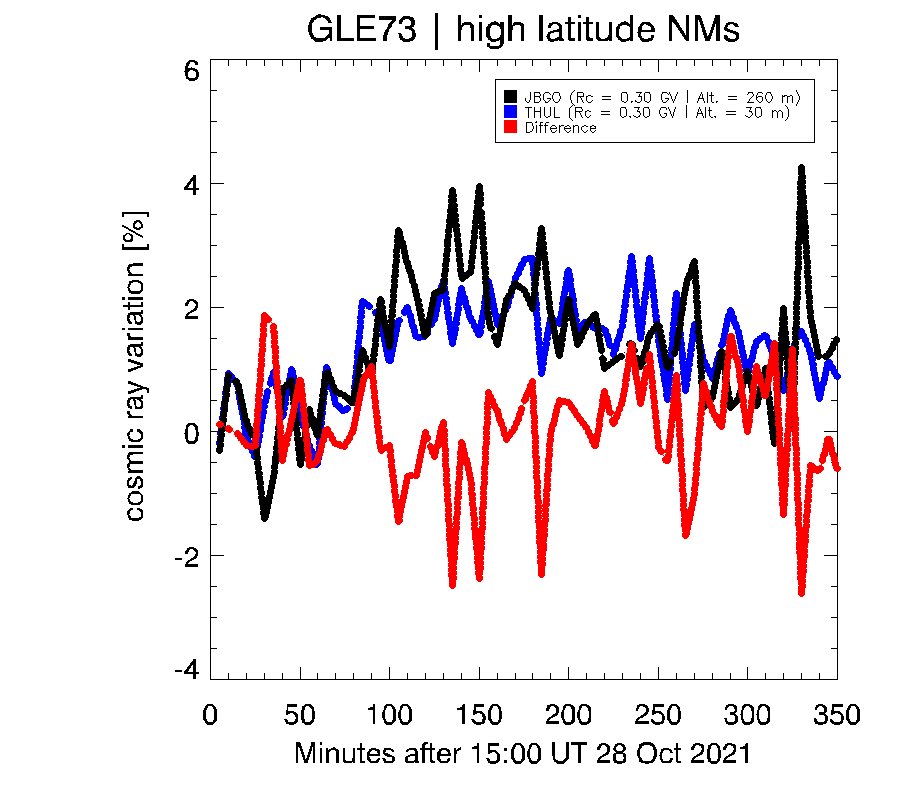}
\caption{Evaluation of north-south anisotropy of SEPs during GLE73. CR variations using two sub-polar stations THUL (North, blue line) and JBGO (South, black line) and their corresponding difference (red line).} 
\label{fig:aniso}
\end{figure}

Moreover, the anisotropy is usually assessed by a direct comparison of count rates of northern and southern near-polar NMs. For GLE73 we compared the count rate of two sub-polar NMs, namely Thule (THUL) and Jang Bogo (JBGO), both of which have similar characteristics (i.e. R = 0.30 GV and altitude of 260 m \& 30 m, respectively). This comparison directly indicates the presence (or not) of north-south anisotropy. As it can be seen in Figure \ref{fig:aniso} the difference (red line) remained close to 0\% (mean = 0.02\%, median = 0.06\%) during GLE73 and thus the comparison does not reveal any significant north-south anisotropy component.

\begin{figure}[h!]
\centering
\includegraphics[width=\columnwidth]{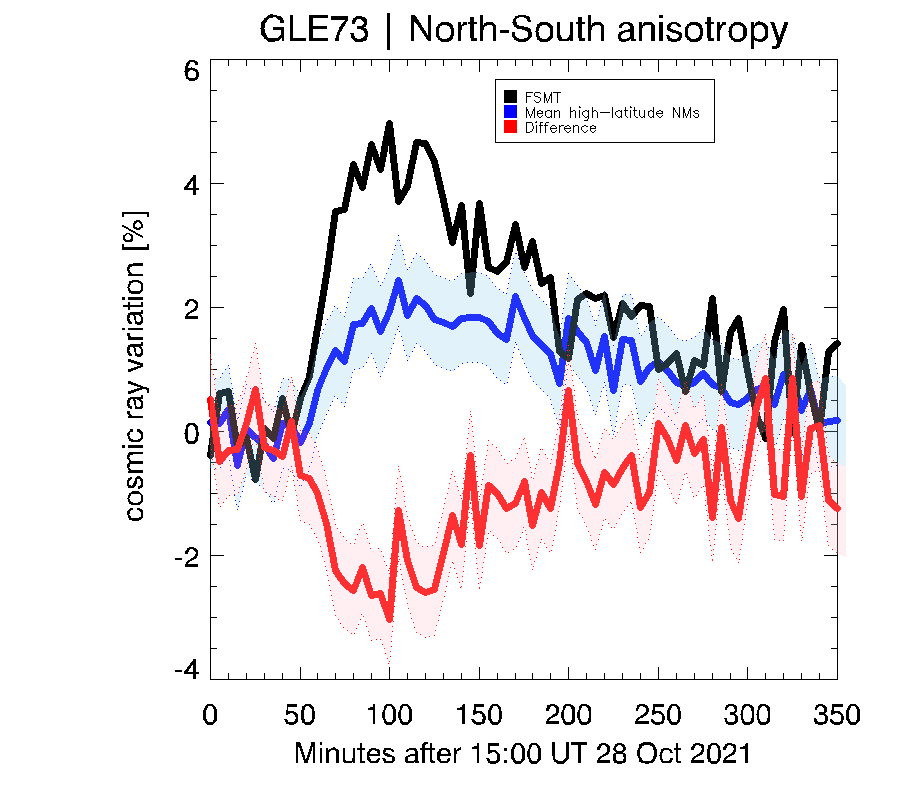}
\caption{Evaluation of high latitude NMs during GLE73. CR variations using FSMT NM (black solid line) and the mean of 8 high latitude NMs (blue solid line) and their corresponding difference (red line). The blue and the pink ribbons depict the 1-$\sigma$ error.} 
\label{fig:figaniso}
\end{figure}

Additionally, in case of an isotropic GLE, high latitude NMs situated at altitudes near the sea level should display almost the same increases caused by SEPs. There are 11 such NM stations with a nominal cut-off rigidity $R_{C}<$1.4 GV \citep{kurt2019onset}.  Figure \ref{fig:figaniso} shows a comparison of the averaged data of 8 of these stations (blue line) against the recordings of the FSMT NM (black line). The fact that FSMT is the only NM which shows a larger increase compared to all other high latitude stations indicates a moderate longitudinal anisotropy of GLE73 in the first $\sim$2 hrs of the event. As it can be seen in Figure \ref{fig:figaniso} the difference (red line) had a maximum of $\sim$3($\pm$0.72)\% during GLE73 at around $\sim$16:40-16:50~UT (mean = 0.92\%, median = 0.87\%).

Table \ref{tab:table1} provides the characteristics (onset, peak time and maximum increase (\%)) of GLE73. Column 1 provides the name (conventional acronym) of the NM used in the analysis, column 2 the GLE onset time (in UT), column 3 the peak time (also in UT), and column 4 the maximum increase (in \%) of the NM station. All products were calculated based on 5-min de-trended NM data \citep{usoskin2020revised}. Although, finer time resolution data (i.e. 1-min) would in principal facilitate a better relation to the solar source, the statistical fluctuations for such a moderate GLE, however, would be too large. DOMC and SOPO as high-altitude and high-latitude stations with a vertical cut-off rigidity of 0.10 GV, allow the registration of lower energy particles compared to the bulk of NMs, that is, they are more sensitive \citep{kuwabara2006real,Mishev2021AP}. As a result, these NMs recorded the most intense flux during GLE73 compared to all other NMs. At the same time the bare NMs at these locations (i.e. DOMB \& SOPB) recorded the most pronounced signals of solar particles for GLE73 (see Table \ref{tab:table1}).\\

\begin{table}[h!]
\centering
\caption[]{Characteristics of GLE73 as recorded by Neutron Monitors (NMs). }
\label{tab:table1}
\begin{tabular}{r|cccc}
\hline 
\hline
Neutron & $R_{c}\dagger$ & Onset & Peak & Increase\\
 Monitor & (GV) & Time (UT) & Time (UT) & (\%)\\
\hline
DOMB  & 0.01 & 16:00$\ast$ & 18:15 & 14.0 \\
SOPB  & 0.09 & 15:50 & 16:30 & 6.64 \\
\hline
DOMC  & 0.01 & 16:00$\ast$ & 18:10 & 7.30 \\
SOPO  & 0.09 & 15:45 & 17:00 & 5.40 \\
PWNK  & 0.16 & 15:55 & 16:20 & 5.10\\
FSMT  & 0.38& 15:50 & 16:40 & 4.97 \\
CALG  & 1.08 & 15:45 & 16:05 & 5.01\\    
SNAE  & 0.56 & 16:15$\ast$ & 17:20 & 4.86 \\
KERG  & 1.01 & 16:05 & 16:50 & 4.15 \\
INVK  & 0.16 & 16:05 & 17:55 & 3.55 \\
TERA  & 0.02 & 16:20$\ast$ & 17:50 & 3.28 \\
OULU  & 0.69 & 15:55$\ast$ & 17:00 & 3.24 \\
YKTK  & 1.65$\star$ & 16:05$\ast$ & 16:35 & 3.10 \\
THUL  & 0.10 & 16:15$\ast$ & 18:55 & 2.83 \\
\hline
 \multicolumn{5}{p{\columnwidth}}{\textbf{Notes}. The de-trended NM data from the International GLE (IGLE) database are being used. The top two rows refer to the bare NMs while the rest to the conventional NMs.}\\
\multicolumn{5}{l}{$\dagger$ from \cite{mishev2020current}}\\
\multicolumn{5}{l}{$\star$ from Yakutsk NM}\\
\multicolumn{5}{l}{$\ast$ ambiguous due to data fluctuations}
\end{tabular}%
\end{table}

\begin{table*}[htp]
\caption{Derived spectral and angular characteristics during several stages of GLE 73 on 28 October 2021.}  \label{table:spectra}      
\centering   
\tiny
\begin{tabular}{c c c c c c c c}     
    

\hline   
\hline    
Integration interval  & $J_{0}$  & $\gamma$ & $\delta\gamma$ & $\sigma$  & $\Psi$ & $\Lambda$  & $\mathcal{D}$ \\ 
 \small[UT\small]           &  [m$^{-2}$ s$^{-1}$ sr$^{-1}$ GV$^{-1}$] &  &  &  [rad $^{2}$] & [degrees] &  [degrees] & [$\%$]\\ 
\hline                    
16:00--16:05 & 77000  & 4.2    &  1.2     & 2.8   & -12.0   & -127  & 21.0 \\  
16:15--16:20 & 82000  & 4.3    &  1.8     & 2.9   & -15.0   & -131  & 14.0 \\  
16:30--16:35 & 94000  & 4.5    &  0.8     & 3.1   & -30.0   & -132  & 11.0 \\  
16:45--16:50 & 96700  & 4.9    &  0.6     & 3.5   & -38.0   & -140  & 12.0 \\  
17:00--17:05 & 103800  & 5.5    &  0.4     & 4.2   & -49.0   & -148  & 10.0 \\  
17:15--17:20 & 108000  & 5.6    &  0.37     & 4.2   & -50.0   & -148  & 11.0 \\  
17:30--17:35 & 114000  & 5.8    &  0.35    & 4.5   & -57.0   & -151  & 10.0 \\  
17:45--17:50 & 119200  & 6.1    &  0.3     & 5.1   & -55.0   & -153  & 12.0 \\  
18:00--18:05 & 125000  & 6.3    &  0.3     & 5.3   & -58.0   & -155  & 11.0 \\  
18:15--18:20 & 131400  & 6.5    &  0.2     & 5.9   & -55.0   & -158  & 14.0 \\  
18:30--18:35 & 115000  & 6.9    &  0.2     & 6.5   & -42.0   & -162  & 11.0 \\  
\hline   
 \multicolumn{8}{p{\columnwidth}}{\textbf{Notes}.  Spectra was modelled with a modified power law rigidity spectrum (Eq.(\ref{eq:eq1})) and PAD with a single Gaussian (Eq.(\ref{eq:eq2})).}\\
\end{tabular}
\end{table*}

    



\noindent \underline{\textit{\textbf{Calculation of the asymptotic directions}}}
A straightforward computation of the rigidity cut-offs and asymptotic directions of the allowed trajectories \citep{cooke1991cosmic} requires a combination of the International Geomagnetic Reference Field (IGRF) geomagnetic model \citep{Alken2021} for the internal field model  with the Tsyganenko 89 model for the external field \citep{tsyganenko1989magnetospheric}. All computations of the particle transport in the geomagnetic field were performed with the MAGNETOCOSMICS code \citep{magneto}.  It is plausible to assume that the first nearly relativistic protons arriving in the vicinity of the Earth  propagate along the interplanetary magnetic field (IMF). Therefore, a NM whose asymptotic cone is aligned nearly to the IMF is expected to register the earliest signal over the background, that is the event onset, and possibly the greatest count rate increase \citep{bombardieri2008improved, papaioannou2014first}. \\


\noindent \underline{\textbf{Modeling the response of neutron monitors}}  
In the model employed in this work, a modified power-law rigidity spectrum of SEPs is assumed:

\begin{equation} \label{eq:eq1}
J_{\parallel}(P) = J_{0}(P)^{(\gamma+\delta\gamma(P-1))}    
\end{equation}

\noindent where $J_{\parallel}(P)$ is the particle flux arriving from the Sun along the symmetry axis, whose direction is defined by the geographic coordinate angles $\Lambda$ and $\psi$, $\gamma$ is the power-law spectral exponent at a rigidity P = 1 GV, and $\delta\gamma$ is the rate of the spectrum steepening. The pitch angle distribution (PAD) is assumed to be similar to a Gaussian:

\begin{equation}\label{eq:eq2}
G(\alpha(P)) \sim \exp(-\alpha^{2}/\sigma^{2})    
\end{equation}

\noindent where $\alpha$ is the pitch angle, and $\sigma$ is the parameter that corresponds to the width of the pitch angle distribution. The pitch angle is defined as the angle between the asymptotic direction and the axis of anisotropy. Note that a steady convergence and reliable solution are usually obtained when the merit function  $\mathcal{D}$, that is the residual according to \citet{Mishev2021} is $\sim$ 5, yet for weak events it can be about 12--15 \citep[e.g. for details see][and the discussion therein]{vashenyuk2006some, Mishev2021} (see Table \ref{table:spectra}). For the GCR spectrum we employed a parametrisation based on the force-field model \citep{gleeson1968solar}, the full details are given in \cite{usoskin2005heliospheric}, where the local interstellar spectrum (LIS) is considered according to \citet{Vos2015}. The modulation is considered following the procedure by \citet{UsoskinJGR2017}. Here, the modelling of the NM response is performed with a new altitude-dependent NM yield function \citep{MishevNMYFJGR2020}, that is, each NM is modelled with a yield function corresponding to the exact station altitude, leading to significant improvement of the unfolding procedure compared to previous studies  \citep [\textit{e.g}.][]{Cramp97, Mishev2021}. Here we rescaled the DOMC/DOMB mini NMs to a standard 6NM64  similarly to \cite{Caballero-Lopez20167461, Lara20161441}. 

\end{appendix}

\end{document}